\documentclass[%
 reprint,
 amsmath,amssymb,
 aps,
pra,
]{revtex4-2}

\usepackage{graphicx}
\usepackage{dcolumn}
\usepackage{bm}
 \usepackage{latexsym}
 \usepackage{amsmath}
 \usepackage{amsfonts}
 \usepackage{amssymb}
 \usepackage{subfigure}
 \usepackage{color}
 \usepackage{setspace}
 \usepackage{titlesec}
 \usepackage{mathtools}

 \newcommand{\ket}[1]{\ensuremath{|#1\rangle}}
 \newcommand{\bra}[1]{\ensuremath{\langle #1 |}}

 \newcommand{\bc}{\begin{center}}
 \newcommand{\ec}{\end{center}}
 \newcommand{\mf}[1]{\boldsymbol{#1}}
 
 \newcommand{\ii}{i}
 \newcommand{\DP}{\Delta_p}
 \newcommand{\DC}{\Delta_c}
 \newcommand{\DR}{\Delta_r}
 \newcommand{\wP}{\omega_p}
 \newcommand{\wC}{\omega_c}
 \newcommand{\wR}{\omega_r}
 \newcommand{\WP}{\Omega_p}
 \newcommand{\WC}{\Omega_c}
 \newcommand{\WR}{\Omega_r}

\begin{document}

\preprint{APS/123-QED}

\title{Efficient diffraction control using a tunable active-Raman gain medium}

\author{Sandeep Sharma}
 \email{quanta.sandeep@gmail.com}
\affiliation{Department of Physics, Korea Advanced Institute of Science and Technology (KAIST), Daejeon 34141, South Korea}%

\date{\today}

\begin{abstract}
We present a new scheme to create all-optical tunable and lossless waveguide using a controllable coherent Raman process in an atomic rubidium vapor in $\mathcal{N}$-type configuration. We employ a Gaussian Raman field and a Laguerre-Gaussian control field to imprint a high-contrast tunable waveguide-like feature inside the atomic medium. We numerically demonstrate that such a waveguide is able to guide arbitrary modes of a weak probe beam to several Rayleigh length without diffraction and absorption. Our results on all-optical waveguide based scheme may have potential application in lossless image processing, high contrast biomedical imaging and image metrology.
\end{abstract}

\maketitle


\section{\label{sec:level1}Introduction}

Optical diffraction poses a major obstacle in the path of achieving high contrast imaging, efficient information transfer and processing. This is because diffraction deforms the optical beam carrying the information, thereby causing a loss of it and thus limiting the efficiency of information processing and imaging methods. To control optical diffraction, different techniques based on coherent quantum effects such as Electromagnetically Induced Transparency \cite{TruscottAG,KapoorR,FirstenbergO}, Coherent Population Trapping \cite{ShpaismanH}, Double Dark Resonance \cite{SharmaS1}, and Saturated Absorption \cite{DeyTN1} have been proposed.  
The main idea in these techniques is to make use of the coherent effects and spatially modulate the susceptibility of an atomic medium using a suitable spatially dependent strong field. Such modulation of the susceptibility by a spatial dependent strong field leads to formation of waveguide like feature inside the atomic medium. This induced waveguide can thus control the diffraction of arbitrary modes of optical beam propagating through the medium. Apart from diffraction control, such spatial dependent techniques have also been used to study cloning and transfer of images \cite{VermaON1,DingDS,CaoM,QinL,ShiZ}, generation of structured beams \cite{RadwellN, SharmaS2,HamediHR1,SharmaS3}, vortex beams \cite{HamediHR2} and in localization of atoms \cite{HamediHR3}.

Moreover, all the above mentioned techniques have been used to create waveguide inside an absorptive medium. Hence, optical beam carrying information suffers from large absorption while propagating in the induced waveguide. To overcome this problem, similar spatial dependent techniques have been used to create waveguide inside active Raman gain (ARG) medium, which supports lossless propagation of arbitrary modes of optical beam without diffraction \cite{VudyasetuPK,DeyTN2}. Nonetheless, this waveguide has its own disadvantage, which is inducing high gain in the optical beam, propagation through the ARG medium. Such high gain can create instability in the medium \cite{AgrawalGP1}. So, to control this high gain, an additional strong control field is applied to the ARG medium \cite{AgarwalGS1}. It has been shown that a strong control field significantly reduces gain for the optical beam propagating through the medium. This interesting behavior shown by the control field have been later used to study optical precursors \cite{PengY}, steering and splitting of optical beams \cite{VermaON2}, and tunable atomic grating \cite{ArkhipkinVG}. However, in the context of diffraction control, the effect of this control field on the induced waveguide inside the ARG medium has remained unexplored.

In this paper, we investigate the effect of this additional control field on the induced waveguide inside the ARG medium. In this regard, we first start by creating a waveguide inside the medium using a Gaussian Raman beam. We find that the features of the waveguide such as width of the core and the length of the waveguide can be controlled by the width of the Raman beam. Most importantly, we next use a Laguerre-Gaussian control beam and examine its effect on the properties of the induced waveguide. We find that this control beam induces a waveguide with very narrow core without changing its length, unlike earlier case where waveguide length decreases with narrowing of core. Waveguides with narrow core plays a very crucial role in controlling diffraction of optical beams having very narrow feature sizes. Our numerical result on optical beam propagation confirms that this control beam induced waveguide can efficiently control diffraction of arbitrary modes of optical beams with feature size $\sim 10 \mu$m to several Rayleigh lengths of the beam, while the waveguide induced by only Raman beam fails in this aspect. Further, controlling diffraction of optical beams with narrow feature size serves an importance purpose of achieving high density information processing and high contrast imaging. Hence, in this regards, this work is very important and holds a greater advantage over other absorptive/gain based induced waveguides, as the later ability to guide arbitrary modes of optical beams are limited to a feature size of $\sim 50 \mu$m and above \cite{TruscottAG,ShpaismanH,DeyTN1,DeyTN2}.

This paper is organized in following manner: In Sec.~(\ref{sec:level2}), we propose our model system and present the governing density matrix equations for the system. 
Next we provide an analytical expression for the susceptibility of the medium in the steady state limit and then the beam propagation equation in the paraxial limit. In Sec.~(\ref{sec:level3}), we present our results on creation of Raman and control field induced tunable waveguide and on its ability to control diffraction of arbitrary modes of optical beams. Finally, in Sec.~(\ref{sec:level4}) we conclude our work and suggest its potential application in the field of optics.

\section{\label{sec:level2}Theoretical Formalism}

\subsection{ Model system}

\begin{figure}[t]
\centering
\includegraphics[width=9.0cm, height=5.2cm]{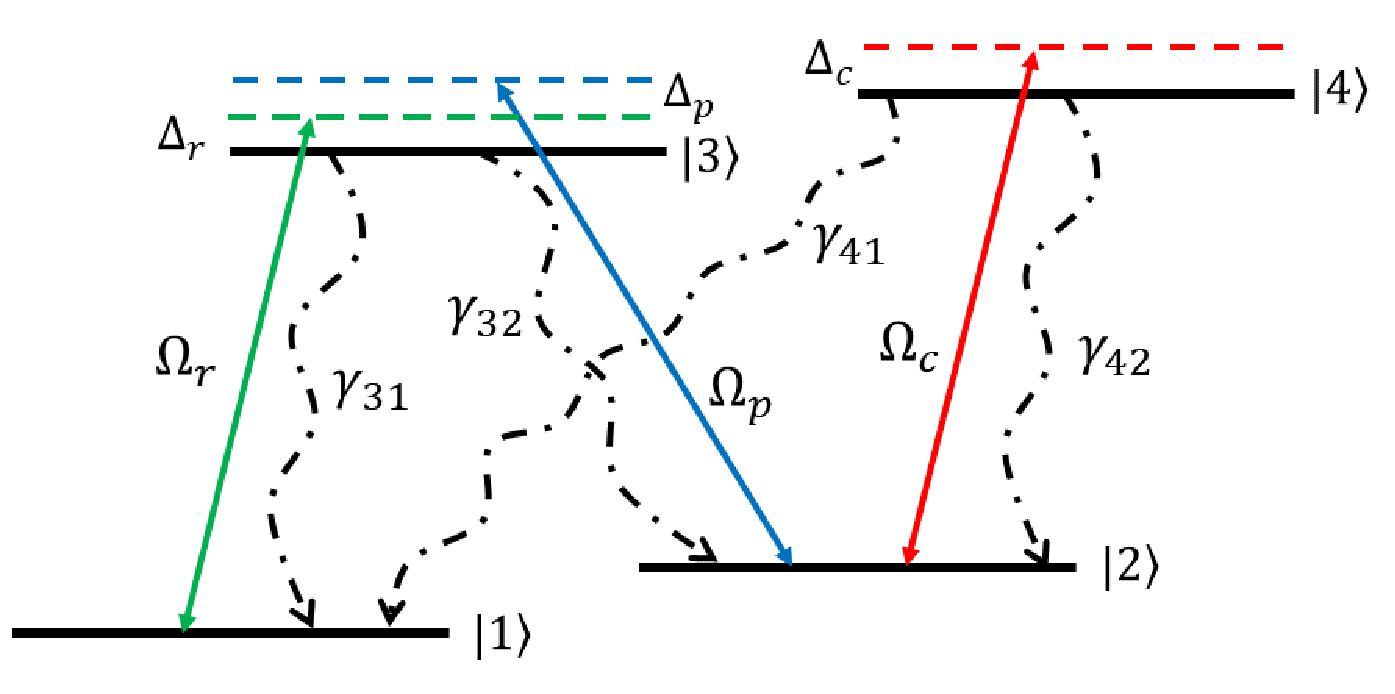}
\caption{\label{fig:Fig1} (Color online) Displays a schematic diagram of four level $\mathcal{N}$-type atomic system. The transitions $\ket{3}\leftrightarrow\ket{1}$ is coupled by a strong Raman field with Rabi frequency $\WR$. A weak probe field with Rabi frequency $\WP$ couples the transition $\ket{3}\leftrightarrow\ket{2}$. The transition $\ket{4}\leftrightarrow\ket{2}$ is driven by a strong control field with Rabi frequency $\WC$. $\gamma_{ij}$ are the spontaneous decay rates from excited states $\ket{i}$ to ground states $\ket{j}$. The field detunings for Raman, probe and control are represented as $\Delta_{r}$, $\Delta_{p}$, and $\Delta_{c}$, respectively. }
\end{figure}

In this work, we consider a tunable ARG system in a $\mathcal{N}$-type configuration as shown in Fig.\ref{fig:Fig1}. The proposed configuration can be experimentally realized by considering the energy levels of $^{87}Rb$, with the two metastable ground states designated as $\ket{1}= \ket{5S_{\frac{1}{2}},F = 2}$, and $\ket{2}= \ket{5S_{\frac{1}{2}},F = 3}$, and the remaining two excited states as $\ket{3}= \ket{5P_{\frac{1}{2}},F^{'} = 2}$ and $\ket{4}= \ket{5P_{\frac{3}{2}},F^{'} = 4}$. As depicted in Fig.\ref{fig:Fig1}, we consider three laser fields acting on the electric-dipole allowed transitions $\ket{3}\leftrightarrow\ket{1}$, $\ket{3}\leftrightarrow\ket{2}$ and $\ket{4}\leftrightarrow\ket{2}$, thereby forming a $\mathcal{N}$-type configuration. The transition $\ket{3}\leftrightarrow\ket{1}$ is driven by a strong Raman field$~(\mathcal{E}_r)$ with frequency $\omega_r$, while a weak probe field$~(\mathcal{E}_p)$ with frequency $\omega_p$ couples the transition $\ket{3}\leftrightarrow\ket{2}$. A strong control field$~(\mathcal{E}_c)$ with frequency $\omega_c$, drives the transitions $\ket{4}\leftrightarrow\ket{2}$.
We define the electric field of all the three laser fields as
\begin{equation}
\label{field}
 {\vec{E}_j}(\vec{r},t)= \hat{e}_{j}\mathcal{E}_{j}(\vec{r})~e^{- i\left(\omega_j t-  k_j z\right )} + {c.c.}\,,
\end{equation}
where, $\hat{e}_{j}$ is the unit polarization vector, $\mathcal{E}_{j}(\vec{r})$ is the slowly varying envelope, $\omega_j$ is the frequency of the laser field and $k_j$ is the wave number, respectively.  The index $j\in \{r,p,c\}$ indicate the Raman, probe and control field, respectively.
In presence of these three laser fields, the time-dependent Hamiltonian of the system under the electric dipole approximation can be expressed as
\begin{subequations}
\label{Hschroed}
\begin{align}
H =& H_0 + H_I\,,\\
H_0 =& \hbar\omega_{12} \ket{2}\bra{2} + \hbar\omega_{13} \ket{3}\bra{3} + \hbar\omega_{14} \ket{4}\bra{4}\,,\\
H_I =& - ( \ket{3}\bra{1} \mf{d}_{31}\cdot\mathcal{E}_re^{- i\left(\omega_r t-  k_r z\right )}  \nonumber \\
 &  + \ket{3}\bra{2} \mf{d}_{32}\cdot\mathcal{E}_pe^{- i\left(\omega_p t-  k_p z\right )}\nonumber\\   
 &  + \ket{4}\bra{2} \mf{d}_{42}\cdot\mathcal{E}_ce^{- i\left(\omega_c t-  k_c z\right )}\,+\,\text{H.c.})\,.
\end{align}
\end{subequations}
Here $\mf{d}_{ij} = \bra{i}\mf{d}\ket{j}$ represents the dipole moment of the transition $\ket{i}\leftrightarrow\ket{j}$.
To remove the time dependencies from the Hamiltonian in Eq.~(\ref{Hschroed}), we make use of the following unitary transformation
\begin{subequations}
\begin{align}
\label{unitary}
W= & e^{-\frac{i}{\hbar}Ut} \,, \\
U= &\hbar\wR\ket{3}\bra{3}+\hbar(\wR-\wP)\ket{2}\bra{2} \\ \nonumber
&+\hbar(\wR-\wP+\wC)\ket{4}\bra{4}\,.
\end{align}
\end{subequations}
and rewrite the Hamiltonian into a time-independent form under the rotating wave approximation as
\begin{align}
\frac{\mathcal{H}_{I}}{\hbar} & =  (\DP-\DR)\ket{2}\bra{2}+(\DP-\DR- \DC)\ket{4}\bra{4}  \notag \\
& -\DR\ket{3}\bra{3}  - \WR\ket{3}\bra{1}-\WP \ket{3}\bra{2}-\WC\ket{4}\bra{2}+\text{H.c.}\,,
 \label{Heff}
\end{align}
where $\Delta_{r} = \omega_{r}-\omega_{31}$, $\Delta_{p} = \omega_{p}-\omega_{32}$, and $\Delta_{c} = \omega_{c}-\omega_{42}$ are the detuning of the Raman, probe and control fields, respectively. 
The Rabi frequencies of the Raman($\WR$), probe($\WP$) and the control fields($\WC$) are defined as
\begin{subequations}
\begin{align}
\label{field}
\WR&=\frac{\vec{d}_{31}\cdot\vec{\mathcal{E}}_{\rm{r}}}{\hbar}e^{ik_r z},\WP=\frac{\vec{d}_{32}\cdot\vec{\mathcal{E}}_{\rm{p}}}{\hbar}e^{ik_p z}, \notag \\ \WC&=\frac{\vec{d}_{42}\cdot\vec{\mathcal{E}}_{\rm{c}}}{\hbar}e^{ik_c z}.\nonumber
\end{align}
\end{subequations}
To study the effect of the three laser fields on the atomic populations and coherences of the system, we use the following Liouville equation 
\begin{eqnarray}
\label{Equationden}
 \frac{\partial\rho}{\partial t} =-\frac{i}{\hbar} [\mathcal{H}_{I},\rho] + \mathcal{L}\rho\,.
\end{eqnarray}
Here, the last term $\mathcal{L}\rho$ is Liouvillian operator which describes the effect of relaxation by radiative decay and is expressed as
\begin{equation}
\mathcal{L}\rho =-\sum\limits_{i=3}^4\sum\limits_{j=1}^2 \frac{\gamma_{ij}}{2}\left(\ket{i}\bra{i}\rho-2\ket{j}\bra{j}\rho_{ii}+\rho\ket{i}\bra{i}\right)\,,
\label{idecay}
\end{equation}
where $\gamma_{ij}$ corresponds to radiative decay rates from excited states $\ket{i}$ to ground states $\ket{j}$.
Next, by incorporating Eq.~(\ref{Heff}) and Eq.~(\ref{idecay}) in Eq.~(\ref{Equationden}), we can write the equation for density matrix elements corresponding to the atomic populations and coherences of the system as
\begin{subequations}
\label{Full_density}
\begin{align}
\dot{\rho}_{11}&= \gamma_{31}\rho_{33}+\gamma_{41}\rho_{44}+ \ii \WR^* \rho_{31}- \ii \WR \rho_{13} \,,\\
\dot{\rho}_{22}&= \gamma_{32}\rho_{33}+\gamma_{42}\rho_{44} + \ii \WP^* \rho_{32}
- \ii \WP \rho_{23}+ \ii \WC^* \rho_{42} \nonumber\\
&- \ii \WC \rho_{24} \,,\\
\dot{\rho}_{33}&= -(\gamma_{31}+\gamma_{32})\rho_{33} + \ii \WR \rho_{13}  - \ii \WR^*\rho_{31}
+ \ii \WP \rho_{23} \nonumber\\ &- \ii \WP^* \rho_{32}\,,\\
\dot{\rho}_{21}&= -\left[\gamma_{c} + \ii (\Delta_p-\Delta_r)\right]\rho_{21} - \ii \WR\rho_{23} + \ii \WP^*\rho_{31} \nonumber\\ 
&+ \ii \WC^*\rho_{41} \,,\\
\dot{\rho}_{31}&= -\left[\frac{\gamma_{31}+\gamma_{32}}{2} - \ii \Delta_r\right]\rho_{31}+ \ii \WP\rho_{21} \nonumber\\
&+ \ii \WR(\rho_{11} - \rho_{33})\,,\\ 
\dot{\rho}_{32}&= -\left[\frac{\gamma_{31}+\gamma_{32}}{2} - \ii \Delta_p\right]\rho_{32}+ \ii \WR\rho_{12} 
- \ii \WC\rho_{34}\nonumber\\ 
&+ \ii \WP(\rho_{22} - \rho_{33})\,,
\end{align}
\begin{align}
\dot{\rho}_{34}&= -\left[\frac{\gamma_{31}+\gamma_{32}+\gamma_{41}+\gamma_{42}}{2} - \ii (\Delta_p-\Delta_c)\right]\rho_{34}\nonumber\\
& + \ii \WR\rho_{14} + \ii \WP\rho_{24}- \ii \WC^*\rho_{32} \,,\\
\dot{\rho}_{41}&= -\left[\frac{\gamma_{41}+\gamma_{42}}{2} + \ii (\Delta_p-\Delta_r-\Delta_c)\right]\rho_{41}\nonumber\\ 
&+ \ii \WC\rho_{21}- \ii \WR\rho_{43}\,,\\
\dot{\rho}_{42}&= -\left[\frac{\gamma_{41}+\gamma_{42}}{2} - \ii \Delta_c\right]\rho_{42}- \ii \WP\rho_{43}\nonumber\\
&+ \ii \WC(\rho_{22} - \rho_{44})\,.
\end{align}
\end{subequations}
where, the time derivatives are denoted by overdots and complex conjugate by $``^{*}"$. The remaining equations for the density matrix elements can be found from the property of conjugate $\dot{\rho}_{ji}^*= \dot{\rho}_{ij}$ and the population conservation condition $\sum\limits_{i=1}^{4} \rho_{ii}=1$.

\subsection{\label{sec:sus}Expression for Linear Susceptibility}

In this section, we provide an analytical expression for the linear susceptibility of the system derived in the steady state limit using a perturbative method. In this method, we assume all the atoms to be in state $\ket{1}$, i.e. $\rho_{11} \approx 1$. Now, as a strong Raman field is applied to $\ket{1}\leftrightarrow\ket{3}$ transition, the condition for atoms to remain in state $\ket{1}$ is made possible by considering the Raman field to be highly detuned from excited state $\ket{3}$. In addition to this, we also consider the weak probe field, which connects the adjacent transition $\ket{3}\leftrightarrow\ket{2}$ to be highly detuned. This arrangement of both the fields turns the system into an active Raman gain system where the probe sees Raman gain around the two-photon resonance condition \cite{AgarwalGS1}. Apart from this, we also apply a strong control field to modulate this induced Raman gain in probe field. Further, it is necessary to mention here that the detuning of this control field has a very negligible effect on the population in state $\ket{1}$. So, for this configuration, we expand the density matrix to first order in probe as it is considered weak, and keep only upto the second order terms corresponding to the Raman field as the Raman process is a second-order process \cite{VermaON2}. With these assumption, the density matrix is then expanded in the following manner:
  \begin{align}
  \label{perturb}
  \rho_{_{ij}}&=\rho_{_{ij}}^{(0)}+\frac{\WR}{\gamma}\rho_{_{ij}}^{(1)}+\frac{\WR^{*}}{\gamma}\rho_{_{ij}}^{(2)}+\frac{\WP}{\gamma}\rho_{_{ij}}^{(3)}+\frac{\WP^{*}}{\gamma}\rho_{_{ij}}^{(4)} \nonumber \\
&+\frac{\WR^{2}}{\gamma^{2}}\rho_{_{ij}}^{(5)}+\frac{\left|\WR\right|^2}{\gamma^{2}}\rho_{_{ij}}^{(6)}+\frac{\WR^{*2}}{\gamma^{2}}\rho_{_{ij}}^{(7)}+\frac{\WR\WP}{\gamma^{2}}\rho_{_{ij}}^{(8)}\nonumber \\
&+\frac{\WR^{*}\WP}{\gamma^{2}}\rho_{_{ij}}^{(9)}+\frac{\WR\WP^{*}}{\gamma^{2}}\rho_{_{ij}}^{(10)}+\frac{\WR^{*}\WP^{*}}{\gamma^{2}}\rho_{_{ij}}^{(11)}+\frac{\WP \left|\WR\right|^2}{\gamma^{3}}\rho_{_{ij}}^{(12)}.
  \end{align}
here $\rho_{_{ij}}^{(0)}$ is the zeroth-order solution determined in the absence of all fields and $\rho_{_{ij}}^{(k)}$ with $k\in \{1-12\}$ is the $k^{th}-$order solution of the density matrix equation. 
We now substitute the above equation in Eq.~(\ref{Full_density}) and equate the coefficients of zeroth-order and each of the $k^{th}-$order terms. As a result, we obtain a total of 13 different sets of coupled density matrix equations corresponding to each of the $\rho_{_{ij}}^{(k)}$ and $\rho_{_{ij}}^{(0)}$ terms.
Next, we solve these 13 different sets of equations for $\rho_{_{ij}}^{(k)}$ and $\rho_{_{ij}}^{(0)}$ in the steady state limit and substitute its expression in Eq.~(\ref{perturb}) to obtain the full expression of the atomic populations and coherences $\rho_{ij}$. In this work, we are interested in studying the effect of medium on the probe field, so we will focus on the atomic coherence $\rho_{32}$ which yields the linear response of the medium. We then write this linear response (susceptibility) $\chi_{p}$ of the medium at probe frequency $\wP$ in terms of the atomic coherence $\rho_{32}$ as
\begin{equation}
\begin{aligned}
\label{Equation7}
\chi_{p}(\omega_p)=& \frac{\mathcal{N}\left|d_{32}\right|^2}{\hbar\WP}\rho_{32} \,, \\
\rho_{32}=& \frac{-\ii~\WP\left|\WR\right|^2 }{(\gamma_{32}-\ii\DP)~\Gamma_{a}+\left|\WC\right|^2}\left[\frac{\Gamma_{a}}{(\frac{\gamma_{31}+\gamma_{32}}{2})^{2}+\DR^{2}} + A\right]\,, \\
A=& \frac{\Gamma_{a}\Gamma_{b}-\left|\WC\right|^2}{\left[(\frac{\gamma_{31}+\gamma_{32}}{2})+\ii\DR\right]\left[\Gamma_{b}\Gamma_{c}+\left|\WC\right|^2\right]}\,.
\end{aligned}
\end{equation}
where $\Gamma_{a}=\frac{(\gamma_{31}+\gamma_{32}+\gamma_{41}+\gamma_{42})}{2}-\ii(\DP-\DC)$, $\Gamma_{b}=\frac{(\gamma_{41}+\gamma_{42})}{2}-\ii(\DR-\DP+\DC)$, and $\Gamma_{c}=\gamma_{c}+\ii(\DR-\DP)$. The susceptibility expression in Eq.~(\ref{Equation7}), clearly depicts its dependence on both Raman and control field intensities. Hence, by suitably manipulating both these fields, we can control the susceptibility of the medium and ultimately control the propagation dynamics of the probe field. In the following section, we present the equation which shows how this susceptibility regulates the propagation of probe beam traveling through the atomic medium.


\subsection{\label{sec:eqn}Propagation Equation}

In order to study the propagation of the probe field through the atomic medium, we use the Maxwell's wave equation under slowly varying envelope and paraxial wave approximation and write it in terms of the Rabi frequency of the probe field into the following form 
\begin{equation}
\label{probe}
\frac{\partial}{\partial z} \WP= \frac{\ii }{2{k_p}} \left(\frac{\partial^{2}}{\partial x^{2}}+\frac{\partial^{2}}{\partial y^{2}}\right)~\WP + 2i{\pi}k_p{\chi_{p}}~\WP \,.
\end{equation}
The first terms on the right hand side of Eq.~(\ref{probe}) leads to diffraction of the probe beam propagating either in free space or through any medium.
The second term on the right hand side of Eq.~(\ref{probe}) represents the effect of the medium on the propagation of probe beam. It is interesting to notice from Eq.~(\ref{probe}) that, if we suitably modulate the medium susceptibility in the transverse direction then we can eventually overcome the effect of diffraction. In the next section, we discuss on the manipulation of the medium susceptibility in the transverse direction and its effect on the probe beam propagation.


\section{\label{sec:level3}Numerical Results and Discussion}


\subsection{\label{sec:hom-sus}Homogeneous linear susceptibility}

\begin{figure}[t!]
 \centering
 \subfigure[][]{%
\label{fig:fig2a}%
\includegraphics[height=4.0cm, width=4.1cm]{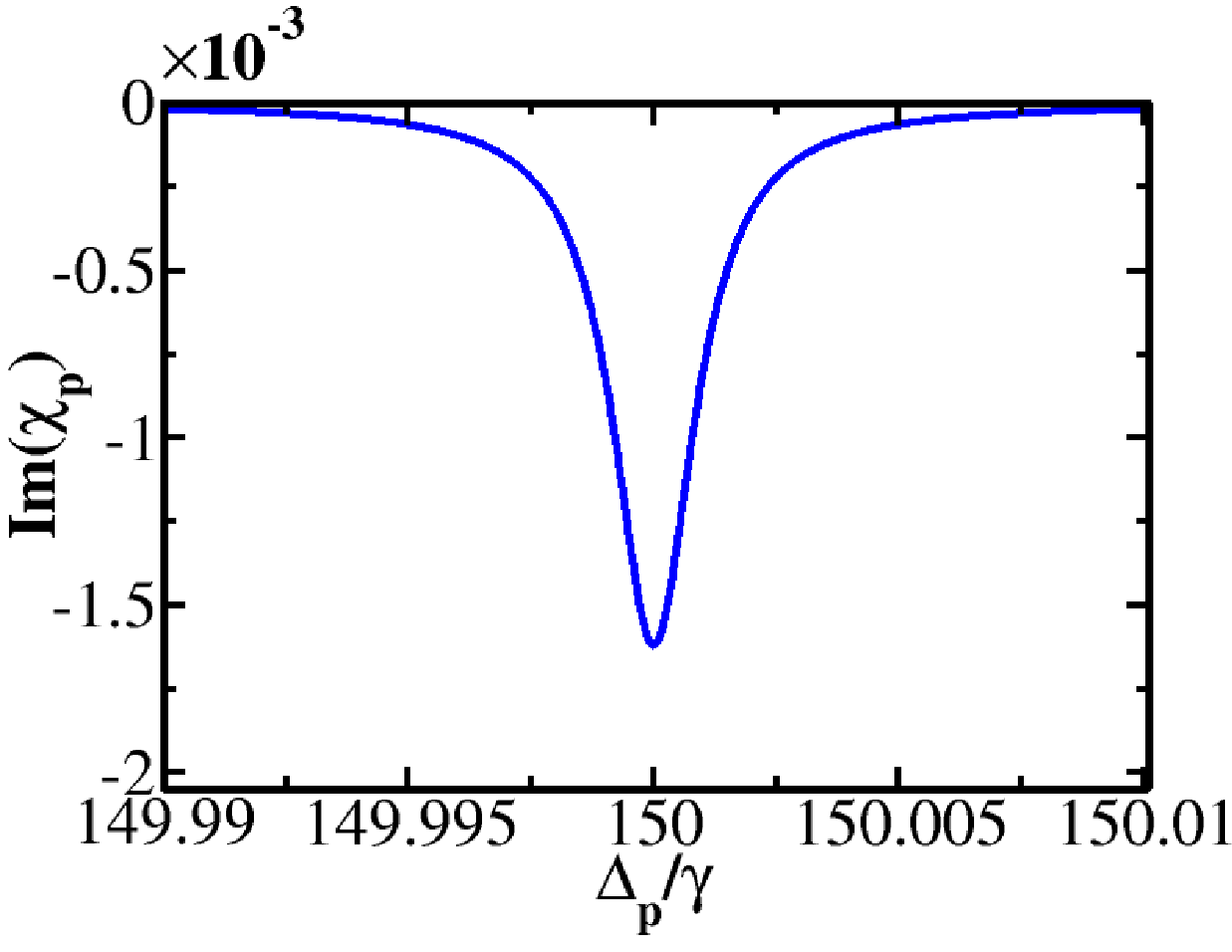}}%
\hspace{8pt}%
\subfigure[][]{%
\label{fig:fig2b}%
\includegraphics[height=4.0cm, width=4.1cm]{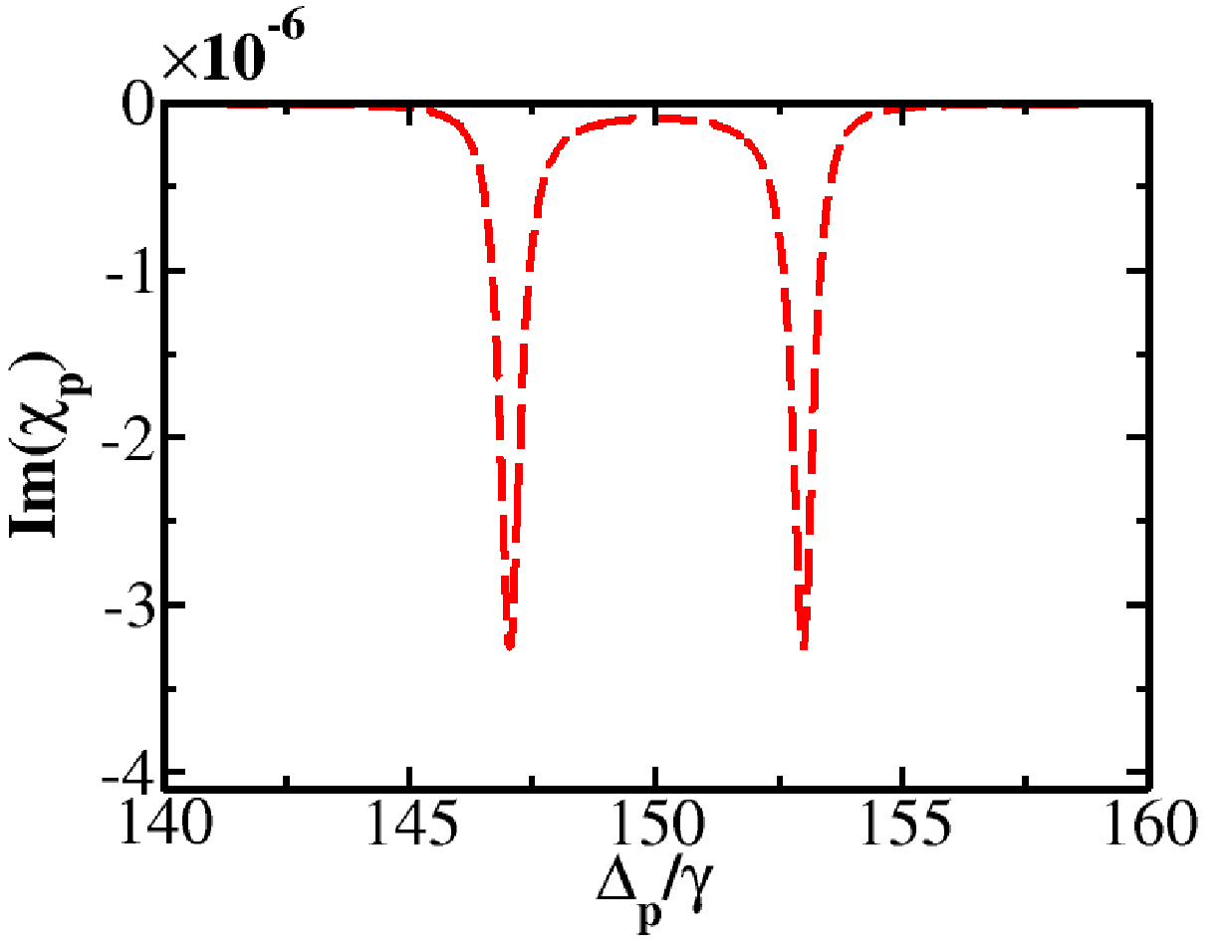}}
 \caption[]{Shows the variation of imaginary part of susceptibility with probe detuning, in absence of control field panel (a) and in presence of control field panel (b). The parameter for panel (a) is $\WR^{0}=2\gamma$ and $\WC^{0}=0\gamma$; for panel (b) is $\WR^{0}=2\gamma$ and $\WC^{0}=3\gamma$. Other parameters are $\gamma_{31}=0.5\gamma$, $\gamma_{32}=0.5\gamma$, $\gamma_{41}=0.5\gamma$, $\gamma_{42}=0.5\gamma$, $\gamma_{c}=10^{-3}\gamma$, $\DR=150\gamma$, $\DC=0\gamma$, $\lambda=795$ nm, and $\mathcal{N}=3\times10^{12}$ atoms/cm$^{3}$.}
 \label{fig:fig2}
 \end{figure}

In this subsection, we first analyze the effect of continuous wave(cw) Raman and control field on the medium susceptibility. In this regard, we numerically evaluate the medium susceptibility as in Eq.~(\ref{Equation7}), first in presence of only cw Raman field and then in presence of both cw Raman and cw control field. The result of our simulation is presented in Fig.~(\ref{fig:fig2}), where the plot shows the variation of imaginary and real parts of susceptibility with probe detuning. In Fig.~\ref{fig:fig2a}, it can be seen that in presence of Raman field the probe sees a Raman gain around the two photon resonance condition $\DP-\DR=0$. This is due to the fact that a strong Raman field acts between the levels $\ket{3}\leftrightarrow\ket{1}$ that has population in lower state and thus induces gain for the weak probe field which acts on the unpopulated levels $\ket{3}\leftrightarrow\ket{2}$, under high detuned condition \cite{AgarwalGS1}. Further, when a strong control field is applied to $\ket{4}\leftrightarrow\ket{2}$ transition, then the probe sees significant reduction in gain at the two photon resonance condition because of the splitting of the single Raman gain peak into two, as shown in Fig.~\ref{fig:fig2b}. This splitting can be attributed to the fact that a strong control field dresses the levels $\ket{4}$ and $\ket{2}$ into the superposition states $\ket{\pm}=(\ket{4}\pm\ket{2})/\sqrt{2}$. As a result, the probe sees two Raman gain peaks corresponding to the transition $\ket{3}\leftrightarrow\ket{+}$ and $\ket{3}\leftrightarrow\ket{-}$, respectively. From above discussion, it is evident that a strong control field plays an important role in controlling the Raman gain seen by the probe. So, in the next section, we discuss on how this Raman gain modulation by the strong control field can be a major factor in creation of tunable waveguide like structure inside the atomic medium.

\subsection{\label{sec:inhom-sus}Inhomogeneous linear susceptibility}

In this subsection, we investigate the effect of spatially dependent Raman and control field on the susceptibility of the atomic medium. For this, we consider Raman field as a Gaussian beam and the control field as a Laguerre-Gaussian(LG) beam. The spatial profile of both the beams are expressed as 
\begin{subequations}
\begin{align}
\WR(x,y,z)&= \WR^{0}\frac{w_{0r}}{w_{r}(z)}e^{-\left(\frac{r^{2}}{w_{r}^{2}(z)}\right)}e^{\left(\frac{ikr^{2}}{2R_{r}(z)}\right)}\nonumber \\
&e^{\left(kz-i\tan^{-1}\left(\frac{z}{z_{0r}}\right)\right)}\,,\\
\WC(x,y,z)&= \WC^{0}\frac{w_{0c}}{w_{c}(z)}\left(\frac{r\sqrt{2}}{w_{c}(z)}\right)e^{-\left(\frac{r^{2}}{w_{c}^{2}(z)}\right)}e^{\left(\frac{ikr^{2}}{2R_{c}(z)}\right)}\nonumber \\
& e^{i \left(\phi-2\tan^{-1}\left(\frac{z}{z_{0c}}\right)\right)}\,
\end{align}
\end{subequations}
where $r= \sqrt{x^{2}+y^{2}}$ and $\phi=\tan^{-1}\left(\frac{y}{x}\right)$. Also, $w_{j}(z)=w_{0j}\sqrt{1+((z-q)/z_{0j})^{2}}$ and $R_{j}(z)=z+(z^{2}_{0j}/z)$ are the spot size and radius of curvature of the beams, respectively, with $w_{0j}$ being the minimum beam waist at focusing point $q$ and $z_{0j}=\pi w^{2}_{0j}/\lambda$ the Rayleigh length of the beams. The index $j\in \{r,c\}$ which represents the Raman and control beam, respectively. 

\begin{figure}[t!]
 \centering
 \subfigure[][]{%
\label{fig:fig3a}%
\includegraphics[height=3.8cm, width=4.1cm]{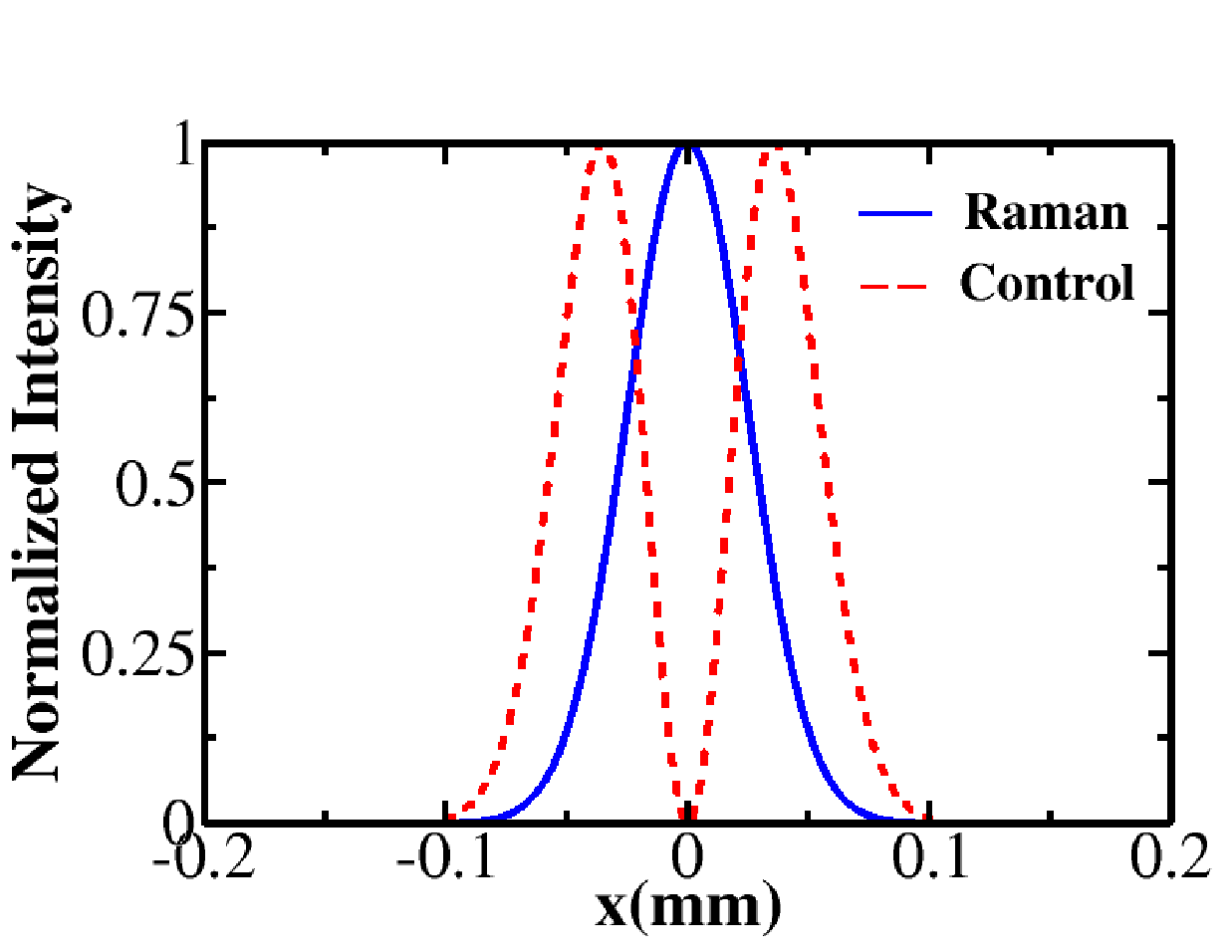}}%
\hspace{8pt}%
\subfigure[][]{%
\label{fig:fig3b}%
\includegraphics[height=3.8cm, width=4.1cm]{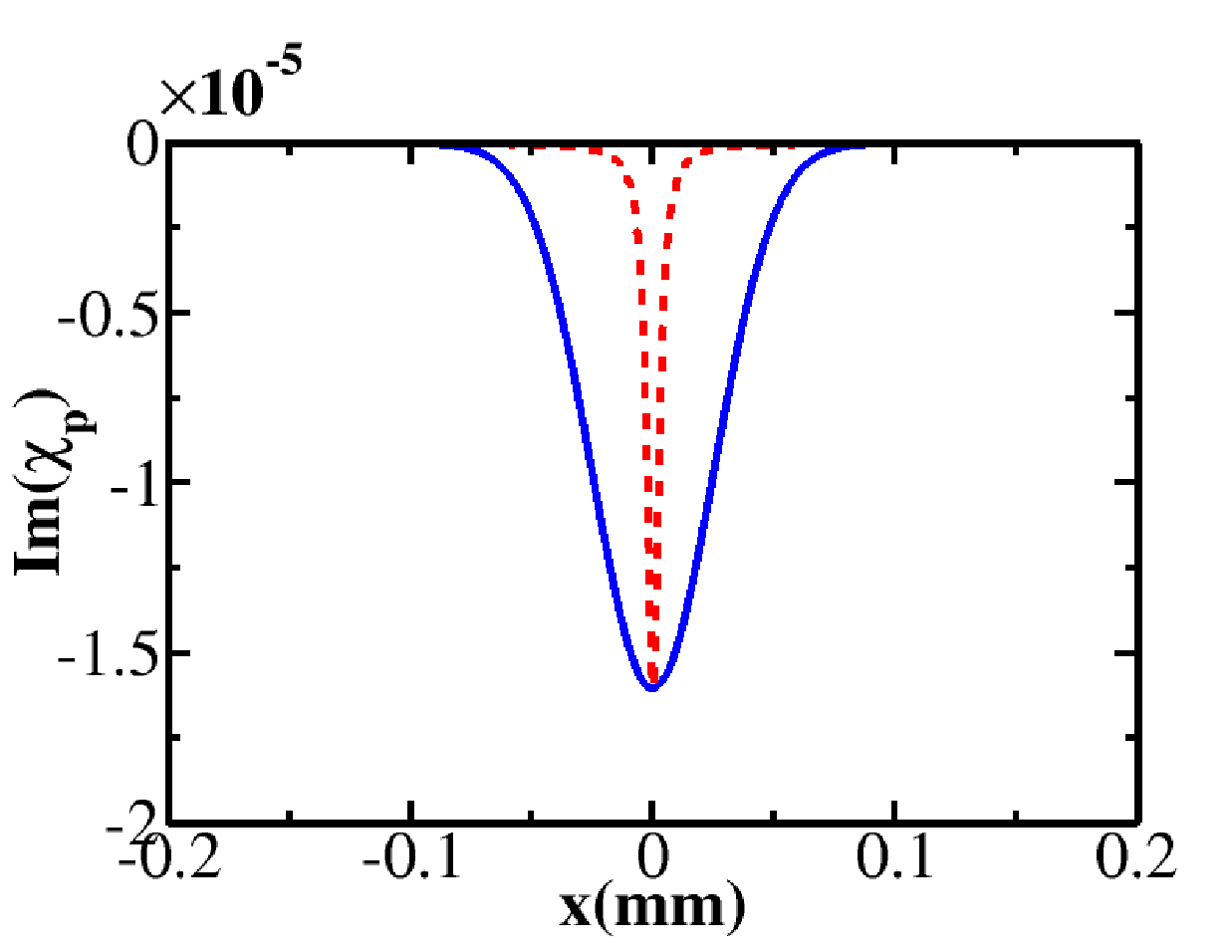}}\\
\hspace{8pt}%
\subfigure[][]{%
\label{fig:fig3c}%
\includegraphics[height=3.8cm, width=4.1cm]{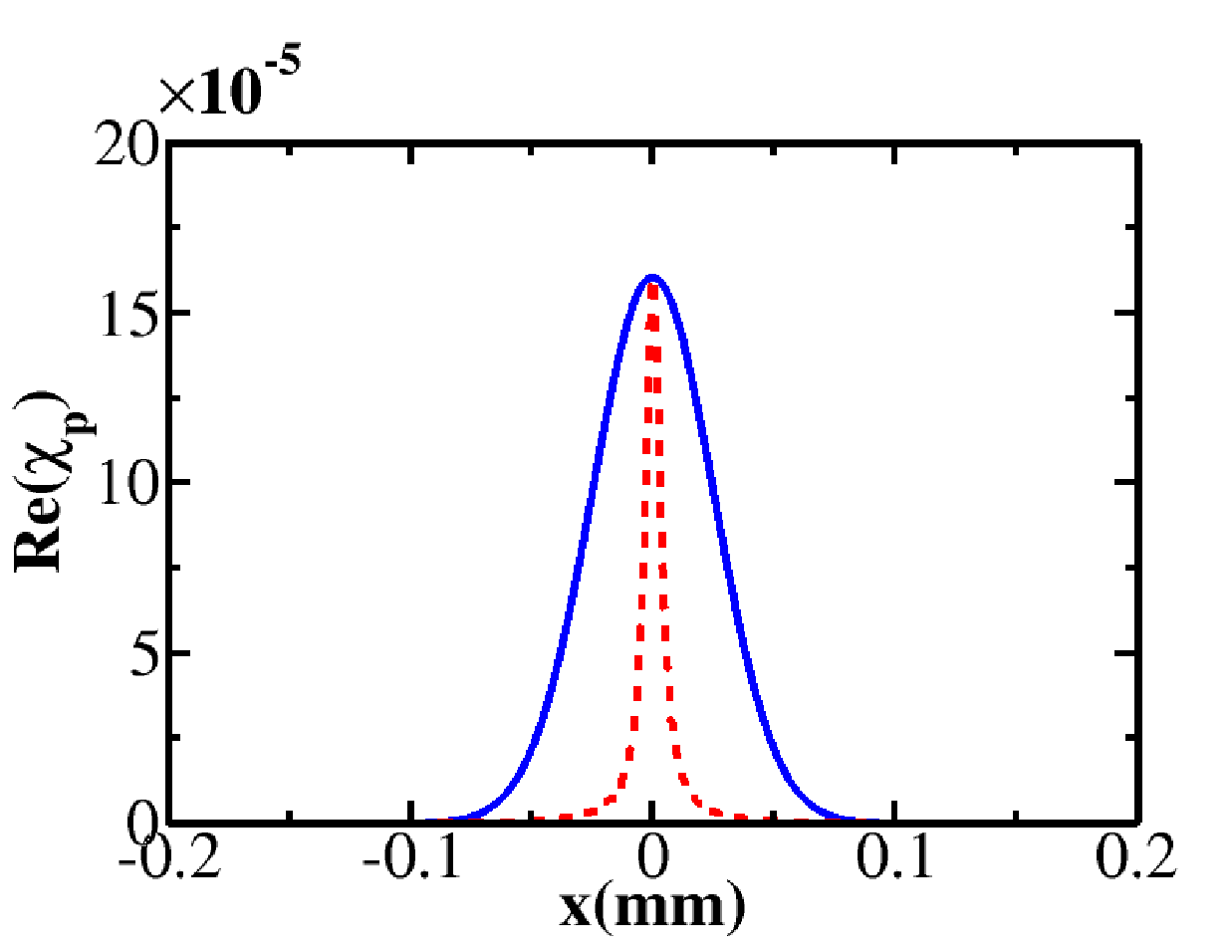}}
 \caption[]{Panel (a) displays the spatial variation of Raman and control field in {\it x}-direction at $y=0$ plane. Panel (b) and (c) shows the variation of imaginary and real part of susceptibility in {\it x}-direction at $y=0$ plane, respectively, in absence (blue-solid line) and in presence (red-dashed line) of control field. The parameter blue-solid line is $\WR^{0}=2\gamma$ and $\WC^{0}=0\gamma$; for red-dashed line is $\WR^{0}=2\gamma$ and $\WC^{0}=3\gamma$. In addition to $w_{r}=50\mu m$, $w_{c}=50\mu m$, and $\DP=150.01\gamma$, all other parameters are same as in Fig.~\ref{fig:fig2} except for $\DC=10\gamma$.}
 \label{fig:fig3}
 \end{figure}

Initially, we consider the case when only Raman field is present and thus examine the behavior of the susceptibility in the transverse direction using  Eq.~(\ref{Equation7}). We present the result of our simulation in Fig.~(\ref{fig:fig3}). Figure~\ref{fig:fig3b} shows that in presence of a Gaussian Raman beam~[See Fig.~\ref{fig:fig3a} for spatial profile], the probe beam sees a spatial gain profile in the transverse direction {\it x} at $y=0$ plane. This can be explained using the fact that Raman beam induces gain in the probe beam as explained in precedent subsection, also shown in Fig.~\ref{fig:fig2a}. Hence, the probe beam will see gain only at the spatial position where Raman beam is present, leading to the formation of a gain window in the transverse direction similar to the profile of Raman beam. Interestingly, when the probe is slightly blue-detuned from the two photon resonance, it sees a waveguide like feature inside the medium as shown in  Fig.~\ref{fig:fig3c}. The variation of the refractive index in the transverse direction as shown in Fig.~\ref{fig:fig3c} depicts a variation similar to waveguide where the central region has higher index (core) than the regions away from center (cladding). Further, it can be seen in Fig.~\ref{fig:fig3c} that the refractive index profile also has a similar spatial shape in the transverse direction like the Raman Gaussian beam. As a result of this spatial similarity, the width of the core region can be considered to be equal to the width of the Raman beam i.e., $w_{r}(z)$. Accordingly, the length of the waveguide can then be attributed to be equivalent to the Rayleigh length $z_{0r}=\pi w^{2}_{0r}/\lambda$ of the Raman beam. Within this length, the width of the Raman beam doesn't change much and hence the properties of the induced waveguide remains intact. So, from above discussion, it is apparent that a spatially dependent Raman beam plays a major role in creating a waveguide like feature inside the medium, which in turn can be able to support diffraction-less propagation of optical beams \cite{VudyasetuPK,DeyTN2}.

Nevertheless, in the next section we show that this Raman induced optical waveguide(RIW) is unable to control efficiently the diffraction of narrow optical beams. To control diffraction of narrow optical beams, a high contrast waveguide with narrow core region is required. For the case of RIW, a high contrast waveguide with narrow core region can be achieved by increasing the atomic density of the system as well as by narrowing the width $w_{r}(z)$ of the Raman beam. This configuration, however, increases gain significantly and also reduces the length of the waveguide ($z_{0r}$) making it less efficient for diffraction control. 

To overcome this problem, we make use of an additional spatial dependent control beam which can control the features of the waveguide. In this regard, we consider the control beam to be a LG beam as shown in Fig.~\ref{fig:fig3a}. This particular spatial profile of the control beam creates a very narrow gain profile as well as a waveguide with a very narrow core region, which is shown in Fig.~\ref{fig:fig3b} and Fig.~\ref{fig:fig3c}, respectively. Such spatial modulation in the gain as well as refractive index of the medium can be attributed to the fact that the presence of control beam decreases gain in the system, as explained in precedent subsection. Now, since the intensity distribution for the LG beam is a donought shape with zero intensity at the center and maximum intensity in the annular region, the spatial gain will be reduced in this annular region, resulting in the narrowing of the spatial gain profile. Similar reduction in the refractive index in the annular region is noticed due to the presence of LG control beam, which leads to the creation of a waveguide with narrow core. 
\begin{figure}[t!]
 \centering
 \subfigure[][]{%
\label{fig:fig4a}%
\includegraphics[height=3.0cm, width=4.cm]{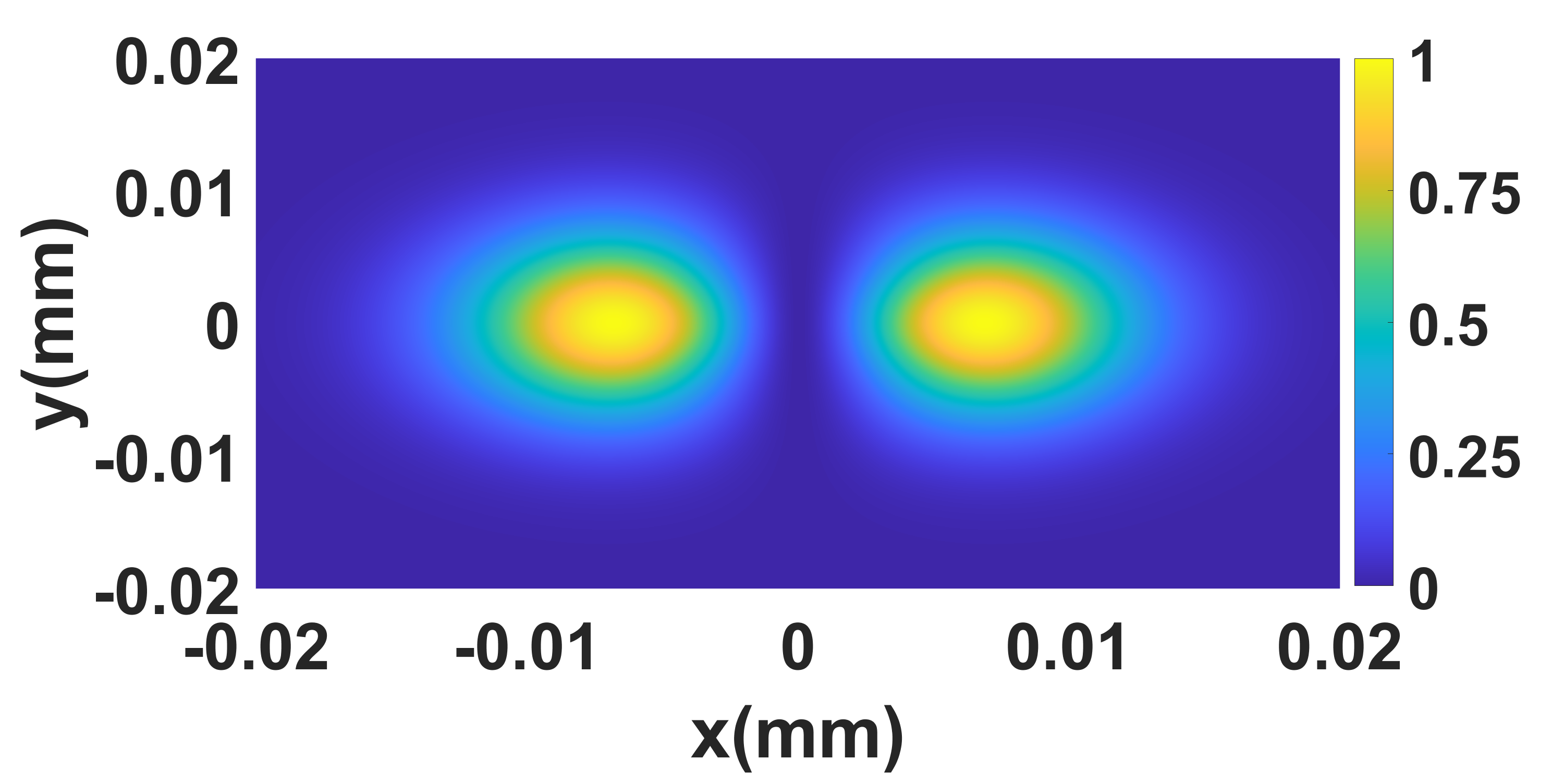}}%
\hspace{8pt}%
\subfigure[][]{%
\label{fig:fig4b}%
\includegraphics[height=3.0cm, width=4.cm]{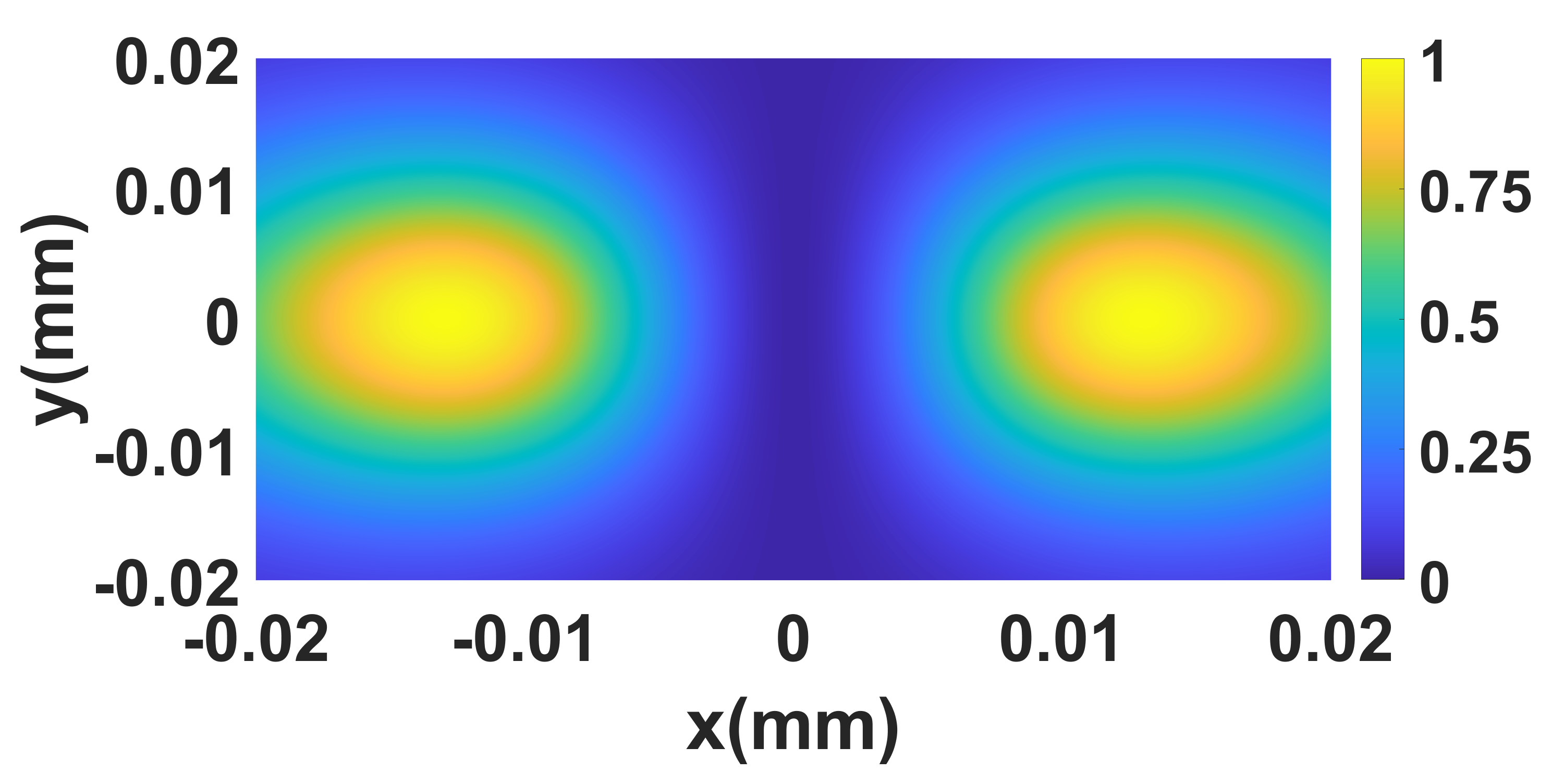}}\\
\hspace{8pt}%
\subfigure[][]{%
\label{fig:fig4c}%
\includegraphics[height=3.0cm, width=4.cm]{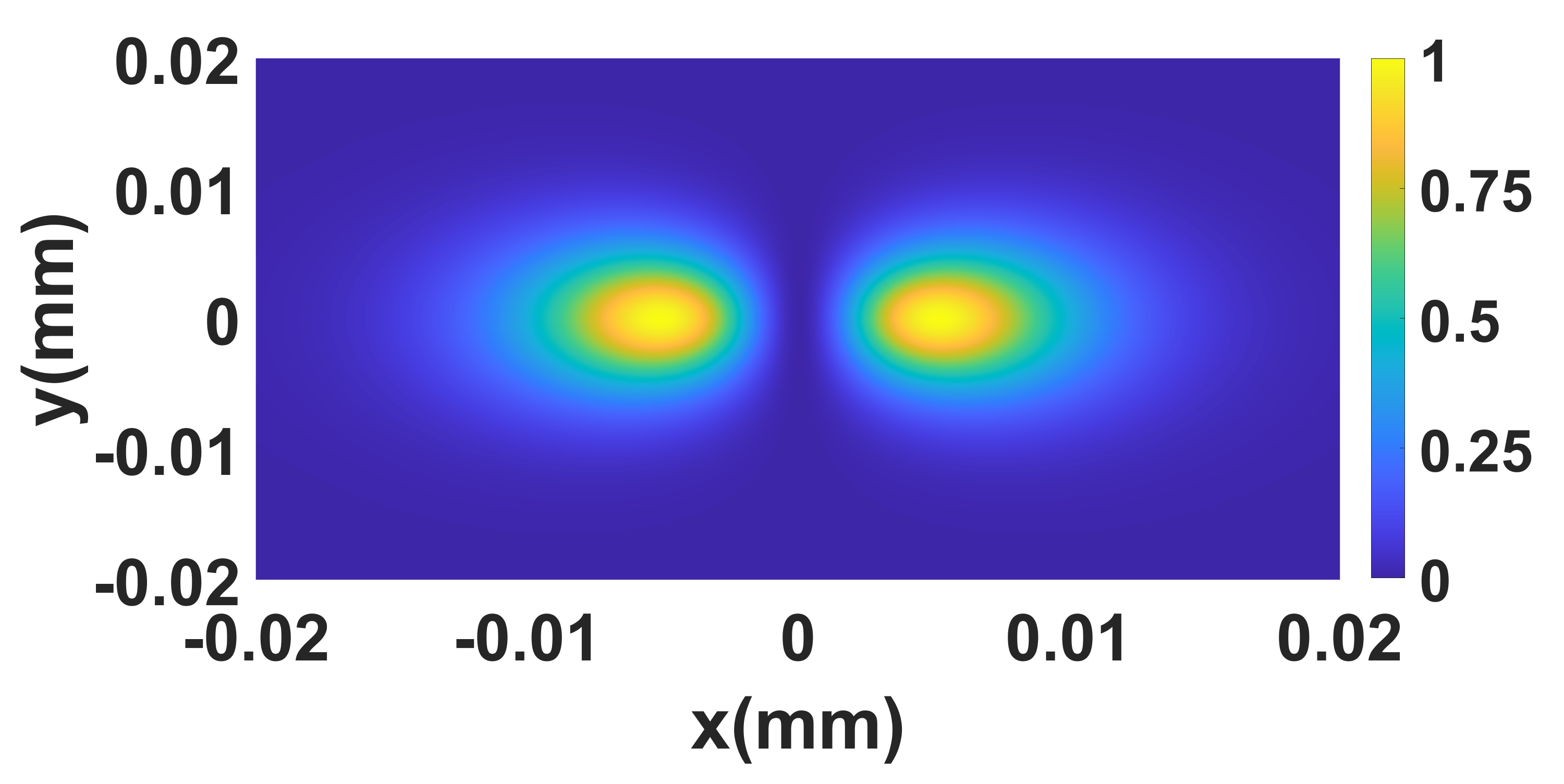}}
\hspace{8pt}%
\subfigure[][]{%
\label{fig:fig4d}%
\includegraphics[height=3.2cm, width=3.9cm]{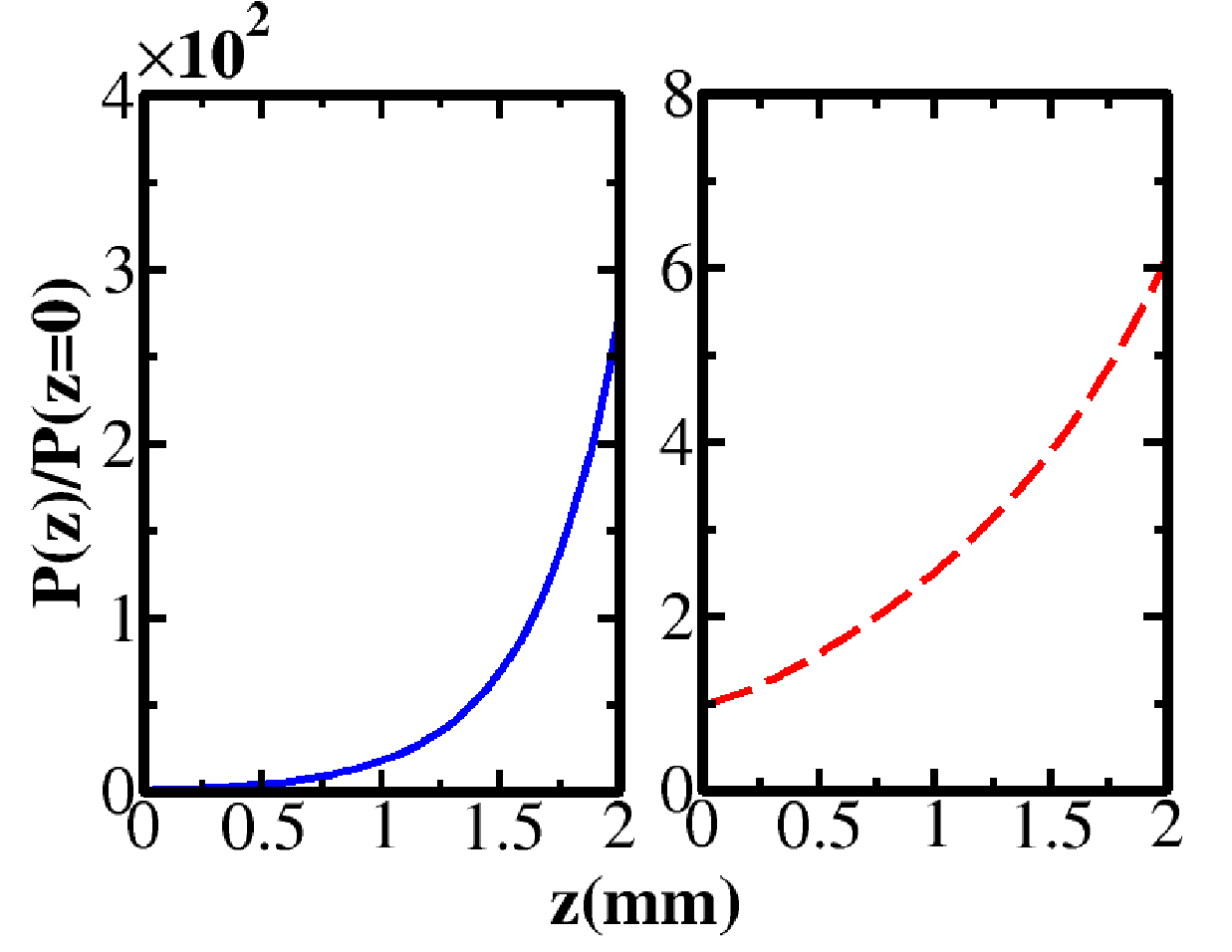}}
 \caption[]{Panel (a) Shows normalized intensity of the input probe beam $(HG_{10})$ in the transverse {\it x-y} plane. Panel (b) and (c) shows the output intensity of probe beam in absence of control beam and in presence of control beam, respectively, at a propagation distance of $z=2$mm. Panel (d) shows the evolution of integrated power of the probe beam in absence of control beam (blue-solid line) and in presence of control beam (red-dashed line), with the propagation distance {\it z}. The parameters are $\WP^{0}=0.001\gamma$, $w_{p}=10\mu m$, $q_{1}=q_{2}=2$, $w_{r}=60\mu m$, $w_{c}=60\mu m$, and $\mathcal{N}=8\times10^{12}$ atoms/cm$^{3}$. Other parameters are same as in Fig.~\ref{fig:fig3}.}
 \label{fig:fig4}
 \end{figure}

\begin{figure}[t!]
 \centering
 \subfigure[][]{%
\label{fig:fig5a}%
\includegraphics[height=3.0cm, width=4.cm]{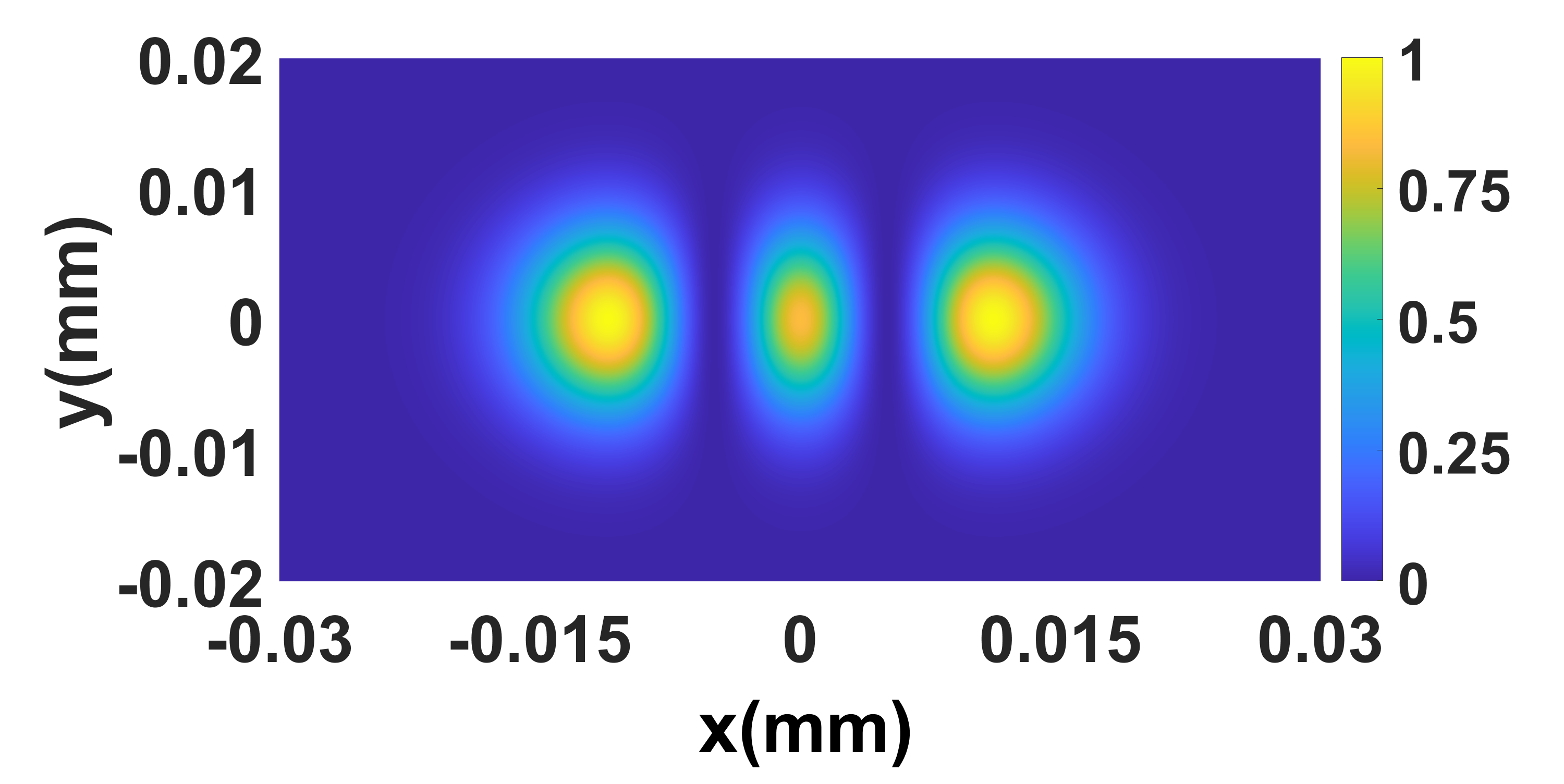}}%
\hspace{8pt}%
\subfigure[][]{%
\label{fig:fig5b}%
\includegraphics[height=3.0cm, width=4.cm]{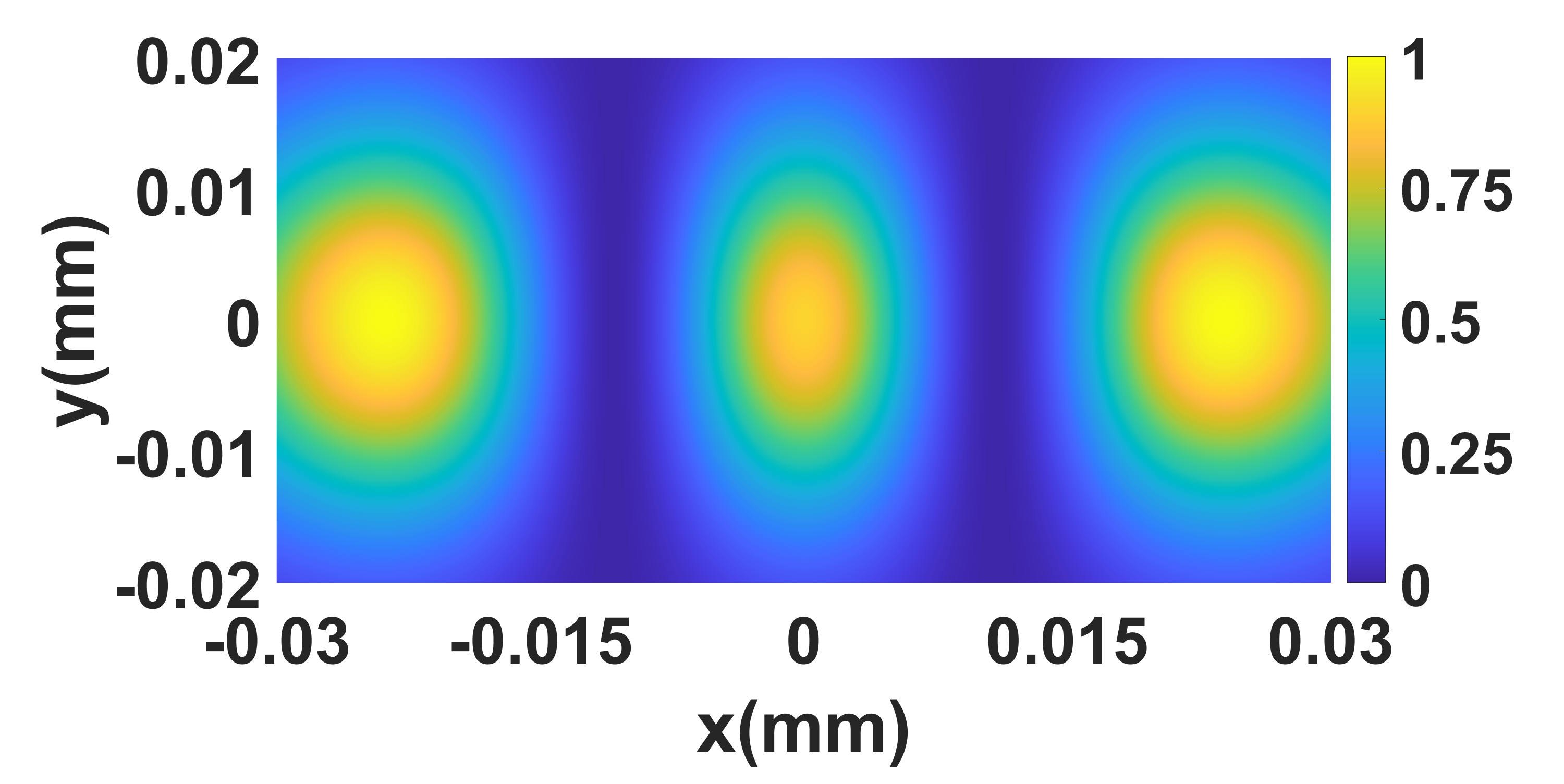}}\\
\hspace{8pt}%
\subfigure[][]{%
\label{fig:fig5c}%
\includegraphics[height=3.0cm, width=4.cm]{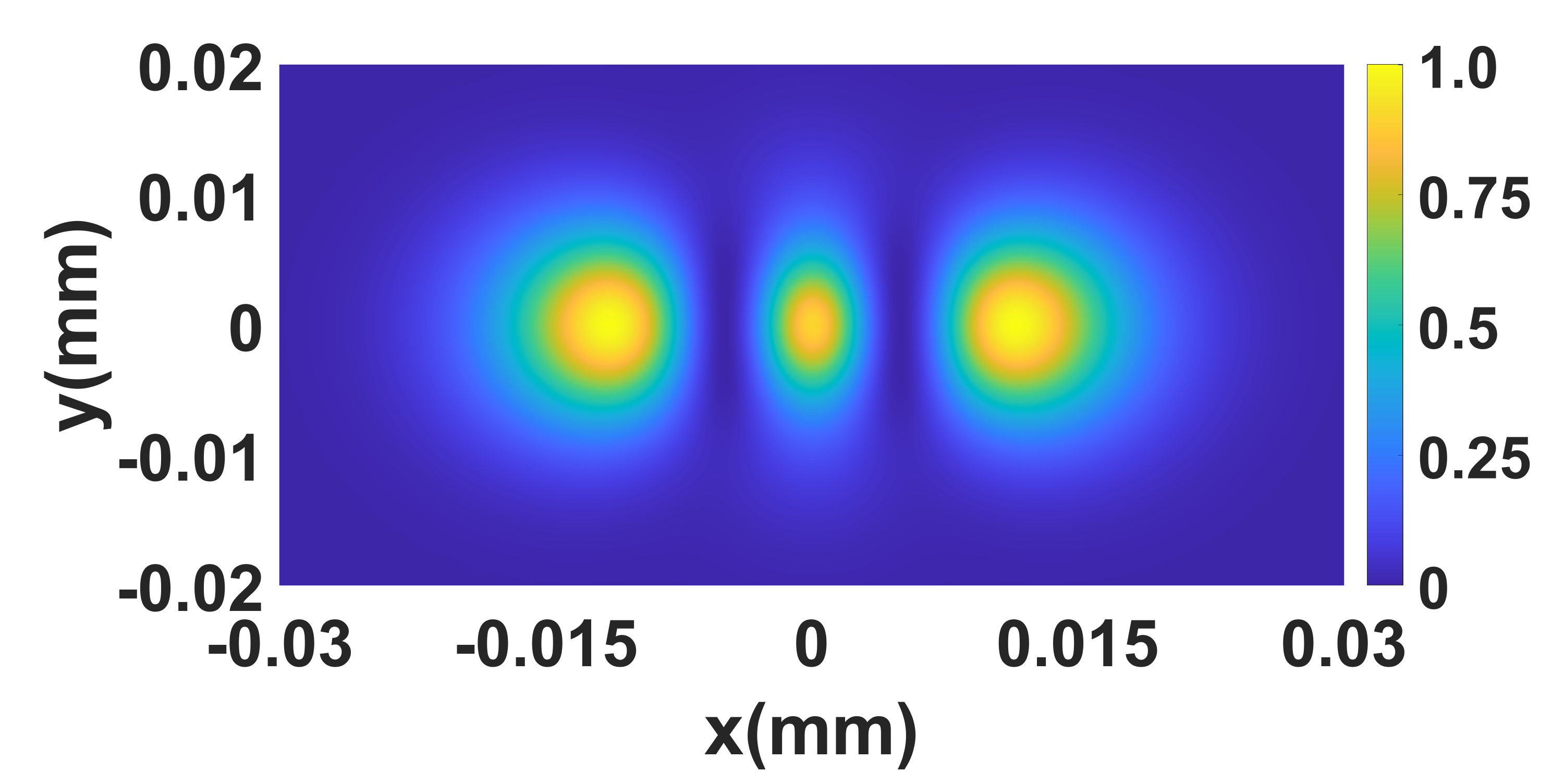}}
\hspace{8pt}%
\subfigure[][]{%
\label{fig:fig5d}%
\includegraphics[height=3.2cm, width=3.9cm]{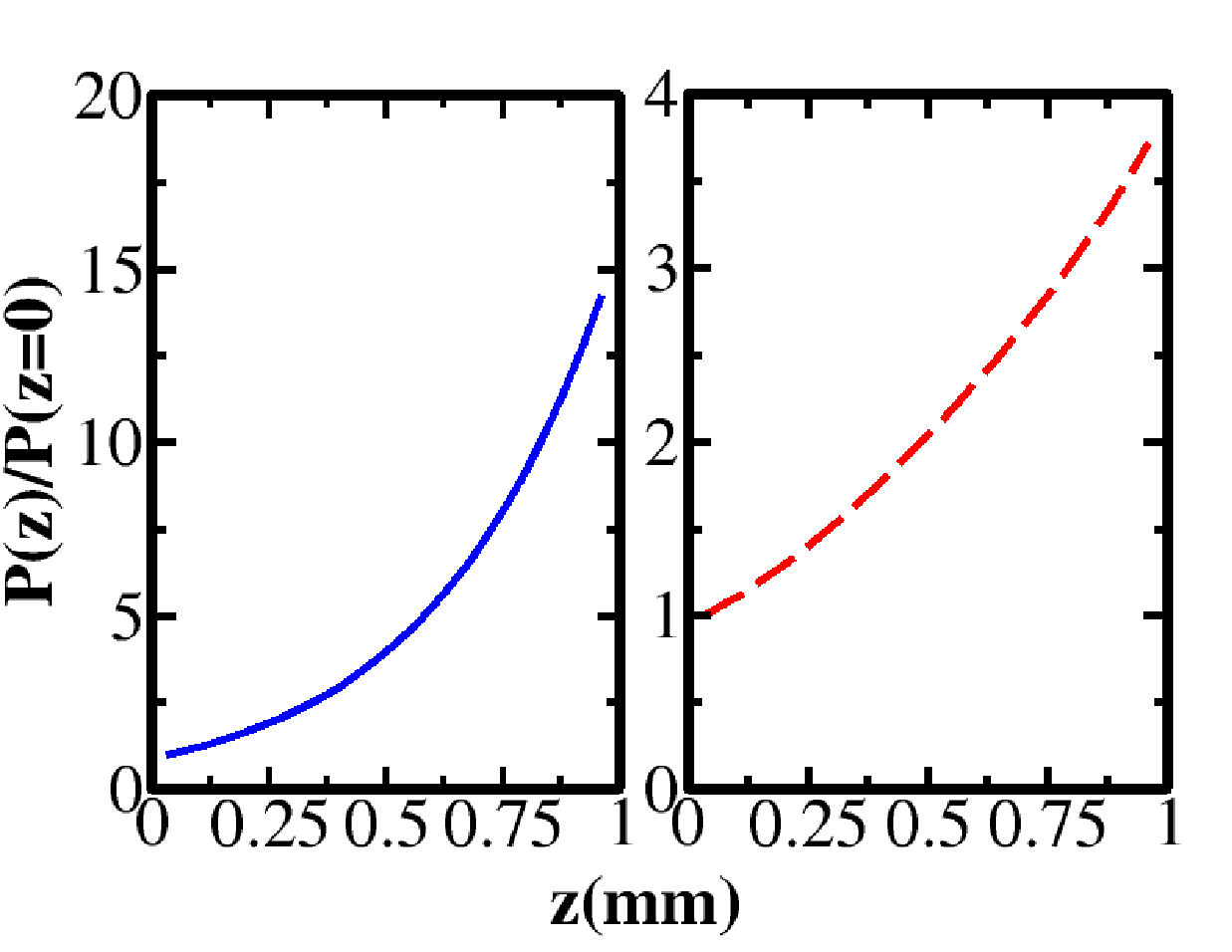}}
 \caption[]{Panel (a) Shows normalized intensity of the input probe beam $(HG_{20})$ in the transverse {\it x-y} plane. The output intensity of probe beam, shown in absence of control beam panel (b) and in presence of control beam panel (c), at a propagation distance of $z=1$mm. Panel (d) shows the variation of integrated power of the probe beam with the propagation distance {\it z}, in absence of control beam (blue-solid line) and in presence of control beam (red-dashed line). The parameters are $w_{r}=90\mu m$, $w_{c}=90\mu m$, $\DR=195\gamma$, $\DC=55\gamma$, $\DP=195.01\gamma$, and $\mathcal{N}=1\times10^{13}$ atoms/cm$^{3}$. Other parameters are same as in Fig.~\ref{fig:fig4}.}
 \label{fig:fig5}
 \end{figure}
Moreover, it is interesting to note that, unlike RIW case where the waveguide narrowing is only possible by narrowing the width of the Raman beam, which leads to a decrease in waveguide length, here the narrowing of waveguide is mainly due to the spatial shape of the control beam, which leaves the waveguide length unchanged. Hence, the spatial profile of the control beam plays a decisive role in controlling the features of the waveguide and in creating high contrast tunable waveguide in the medium. In the following subsection, we show how this control beam tunable Raman induced waveguide(CRIW) holds a greater advantage over RIW in controlling diffraction of narrow optical beams.


\subsection{\label{sec:propagation-sus}Probe beam propagation}

In this subsection, we explore on the possibility of diffractionless propagation of arbitrary modes of narrow optical beams in RIW and CRIW created inside the atomic medium. For this, we consider the initial profile of the probe beam as different Hermite-Gaussian modes, which are expressed as
\begin{equation}
\WP(x,y,z=0)= \WP^{0}H_{n}\left(\frac{\sqrt{2}x}{w_{p}}\right)H_{m}\left(\frac{\sqrt{2}y}{w_{p}}\right)e^{-\frac{\left(x^{2}+y^{2}\right)^{2}}{w_{p}^{2}}},
\end{equation}
where $H_{n}$ and $H_{m}$ are the Hermite polynomials of order $n$ and $m$, respectively. With above mentioned different initial profile, we study the propagation of probe beam by numerically solving Eq.~(\ref{probe}) using Fourier split step method(FSSM) \cite{BandraukAD}. We first consider the propagation of $HG_{10}$ probe beam with a width ($w_{p}$) of 10 $\mu$m in RIW and present the result of our simulation in Fig.~\ref{fig:fig4}. Figure~\ref{fig:fig4b} shows the intensity distribution of the output probe beam in the transverse plane after propagating a distance of $z= 2$mm in RIW. It is clear from Fig.~\ref{fig:fig4b} that probe beam gets diffracted after propagating through this induced wavegide. This is because, RIW has a wide core and is unable to control the diffraction of the probe beam. In addition to this, we also study the output probe power $P(z)=\iint \left|\WP(x,y,z)\right|^2dx~dy$, and found that the probe beam is highly amplified by a factor of 100 after propagating a distance of $z= 2$mm, as shown in Fig.~\ref{fig:fig4d}. So, due to its inability to control diffraction along with creation of such large gain in the probe beam, which can lead to instability in the medium \cite{AgrawalGP1}, the practical applicability of the RIW is severely limited.

However, this problem is eventually overcome by utilizing an additional LG control beam, which can efficiently control both diffraction as well as gain in the probe beam, as shown in Figure~\ref{fig:fig4c} and Figure~\ref{fig:fig4d}, respectively. Such control over diffraction by the control beam can be attributed to the creation of waveguide with narrow core, which is explained in detail in precedent subsection. In addition to diffraction control, the probe beam is also slightly focused as shown in Fig.~\ref{fig:fig4c}. Such focusing behaviour have been observed earlier in atomic based waveguides \cite{DeyTN2} and can be controlled by suitable modulating the features of the waveguide. Further, it is interesting to notice that the probe beam propagates without diffraction in this narrow waveguide [See  Fig.~\ref{fig:fig4c}] upto a distance of $2$mm, which is 5 times the Rayleigh length of the probe beam $z_{0p}$. Also, unlike earlier case, here the probe beam attains a reasonable amount of gain at a distance of $2$mm, as shown in Fig.~\ref{fig:fig4d}. Our result on diffraction control of narrow optical beams to several Rayleigh lengths, may have potential application in high density lossless information transfer to macroscopic lengths and high contrast imaging.

Further, in this work, we are dealing with propagation of optical beams in a waveguide induced inside a gain system. So, it is necessary to verify the validity of both weak field approximation and paraxial approximation used in the calculation. In this regard, for our numerical calculations on beam propagation, we start with low probe power and throughout the propagation we found that the maximum intensity of probe is always very much smaller than the maximum intensity of Raman and Control beam. This condition of probe intensity validates our weak field approximation. Also, the width of the probe beam $w_{p}$ doesn't change much throughout it's propagation and the condition for paraxiality $\lambda/2\pi w_{p}< 0.1$ is always satisfied \cite{AgrawalGP2}, thus validating our paraxial approximation.

Next, to establish that CRIW can support arbitrary modes of the probe beam, we also study the propagation of $HG_{20}$ probe beam in it and present our result in Fig.~\ref{fig:fig5}. From Fig.~\ref{fig:fig5}, it is noticeable that CRIW is able to control the diffraction of the probe beam to length of $2.5 z_{0p}$ with reasonable amount of gain, while RIW fails to do so. Apart from this, we also performed numerical simulation on propagation of other higher order modes of the probe beam and found diffractionless propagation of such modes in CRIW. Hence, from above discussions, it becomes evident that control beam plays an important role in creation of CRIW and subsequently in controlling the diffraction of arbitrary modes of narrow optical beams to several Rayleigh lengths.

\section{\label{sec:level4}Conclusion}
In conclusion, we have investigated diffraction control of arbitrary modes of optical beam using tunable Raman effects arising in atomic medium in ${\mathcal N}$-type configuration. We have demonstrated that by using a suitable spatial dependent Raman field, a waveguide like feature can be created inside the atomic medium. Further, we show that the properties of the induced waveguide can be modulated by using an additional spatial dependent control field. From our numerical results on optical beam propagation, we conclude that RIW is incapable of controlling the diffraction of narrow optical beams, while CRIW can efficiently control the diffraction of such beams to several Rayleigh lengths. Our results may have possible application in high contrast imaging \cite{KapaleKT,LiH}, information processing \cite{DingDS1,DingDS2} and lithography \cite{KiffnerM}.

\begin{acknowledgments}
We wish to acknowledge the financial support from KAIST through the BK21 postdoctoral fellowship.
\end{acknowledgments}

\nocite{*}


\end{document}